\documentclass[]{interact}

\usepackage{epstopdf}
\usepackage[caption=false]{subfig}

\usepackage[ruled,vlined]{algorithm2e}
\usepackage{amsfonts}
\usepackage[hidelinks]{hyperref}
\usepackage{blindtext}
\usepackage{siunitx}
\usepackage{xcolor}
\definecolor{ao}{rgb}{0.0, 0.0, 1.0}
\usepackage[normalem]{ulem}
\usepackage{marginnote}

\usepackage{natbib}
\bibpunct[, ]{(}{)}{;}{a}{}{,}
\renewcommand\bibfont{\fontsize{10}{12}\selectfont}

\renewcommand\cite{\citep}

\theoremstyle{plain}

\theoremstyle{definition}

\theoremstyle{remark}

\DeclareMathOperator{\IM}{\mathbf{DE9IM}}
\DeclareMathOperator{\Crop}{Crop}
\DeclareMathOperator{\Warp}{Warp}

\begin{document}


\title{Deep Homography Estimation in Dynamic Surgical Scenes for Laparoscopic Camera Motion Extraction}

\author{}

\author{
\name{Martin Huber\textsuperscript{1}\thanks{Corresponding author: Martin Huber Email: martin.huber@kcl.ac.uk}, S\'{e}bastien Ourselin\textsuperscript{1}, Christos Bergeles\textsuperscript{1}, and Tom Vercauteren\textsuperscript{1}}
\affil{\textsuperscript{1}School of Biomedical Engineering \& Image Sciences, Faculty of Life Sciences \& Medicine, King's College London, London, United Kingdom}
}

\maketitle 

\begin{abstract}
Current laparoscopic camera motion automation relies on rule-based approaches or only focuses on surgical tools. Imitation Learning (IL) methods could alleviate these shortcomings, but have so far been applied to oversimplified setups. Instead of extracting actions from oversimplified setups, in this work we introduce a method that allows to extract a laparoscope holder's actions from videos of laparoscopic interventions. We synthetically add camera motion to a newly acquired dataset of camera motion free da Vinci surgery image sequences through a novel \textit{homography generation algorithm}. The synthetic camera motion serves as a supervisory signal for camera motion estimation that is invariant to object and tool motion. We perform an extensive evaluation of state-of-the-art (SOTA) Deep Neural Networks (DNNs) across multiple compute regimes, finding our method transfers from our camera motion free da Vinci surgery dataset to videos of laparoscopic interventions, outperforming classical homography estimation approaches in both, precision by $41\%$, and runtime on a CPU by $43\%$. 
\end{abstract}


\section{Introduction}

The goal in IL is to learn an expert policy from a set of expert demonstrations.
IL has been slow to transition to interventional imaging. In particular, the slow transition of modern IL methods into automating laparoscopic camera motion is due to a lack state-action-pair data \cite{kassahun2016surgical, esteva2019guide}. The need for automated laparoscopic camera motion \cite{pandya2014review, ellis2016task} has, therefore, historically sparked research in rule-based approaches that aim to reactively center surgical tools in the field of view \cite{agustinos2014visual, da2020scan}. DNNs could contribute to this work by facilitating SOTA tool segmentations and automated tool tracking \cite{garcia2017toolnet, garcia2021image, gruijthuijsen2021autonomous}.



Recent research contextualizes laparoscopic camera motion with respect to (w.r.t.) the user and the state of the surgery. DNNs could facilitate contextualization, as indicated by research in surgical phase and skill recognition \cite{kitaguchi2020real}. However, current contextualization is achieved through handcrafted rule-based approaches \cite{rivas2014towards, rivas2017smart}, or through stochastic modeling of camera positioning w.r.t. the tools \cite{weede2011intelligent, rivas2019transferring}. While the former do not scale well and are prone to nonlinear interventions, the latter only consider surgical tools. However, clinical evidence suggests camera motion is also caused by the surgeon's desire to observe tissue \cite{ellis2016task}. Non-rule-based, i.e. IL, attempts that consider both, tissue, and tools as source for camera motion are \cite{ji2018learning, su2020multicamera, wagner2021learning}, but they utilize an oversimplified setup, require multiple cameras or tedious annotations.

In current laparoscopic camera motion automation, DNNs merely solve auxiliary tasks. Consequentially, current laparoscopic camera motion automation is rule-based, and disregards tissue. While modern IL approaches could alleviate these issues, clinical data of laparoscopic surgeries remains unusable for IL. Therefore, SOTA IL attempts rely on artificially acquired data \cite{ji2018learning, su2020multicamera, wagner2021learning}. 

In this work, we aim to extract camera motion from videos of laparoscopic interventions, thereby creating state-action-pairs for IL. To this end, we introduce a method that isolates camera motion (actions) from object and tool motion by solely relying on observed images (states). To this end, DNNs are supervisedly trained to estimate camera motion while disregarding object, and tool motion. This is achieved by synthetically adding camera motion via a novel \textit{homography generation algorithm} to a newly acquired dataset of camera motion free da Vinci surgery image sequences. In this way, object, and tool motion reside within the image sequences, and the synthetically added camera motion can be regarded as the only source, and therefore ground truth, for camera motion estimation. Extensive experiments are carried out to identify modern network architectures that perform best at camera motion estimation. The DNNs that are trained in this manner are found to generalize well across domains, in that they transfer to vast laparoscopic datasets. They are further found to outperform classical camera motion estimators.

\section{Related Work}

Supervised deep homography estimation was first introduced in \cite{detone2016deep} and got improved through a hierarchical homography estimation in \cite{erlik2017homography}. It got adopted in the medical field in \cite{bano2020deep}. All three approaches generate a limited set of homographies, only train on static images, and use non-SOTA VGG-based network architectures \cite{simonyan2014very}.

Unsupervised deep homography estimation has the advantage to be applicable to unlabelled data, e.g. videos. It was first introduced in \cite{nguyen2018unsupervised}, and got applied to endoscopy in \cite{gomes2019unsupervised}. The loss in image space, however, can't account for object motion, and only static scenes are considered in their works. Consequentially, recent work seeks to isolate object motion from camera motion through unsupervised incentives. Closest to our work are \citet{le2020deep}, where the authors generate a dataset of camera motion free image sequences. However, duo to tool, and object motion, their data generation method is not applicable to laparoscopic videos, since it relies on motion free image borders. \citet{zhang2020content} provide the first work that does not need a synthetically generated dataset. Their method works fully unsupervised, but constraining what the network minimizes, is difficult to achieve.

Only \cite{le2020deep} and \cite{zhang2020content} train DNNs on object motion invariant homography estimation. Contrary to their works, we train DNNs supervisedly. We do so by applying the data generation of \citet{detone2016deep} to image sequences rather than single images. We further improve their method by introducing a novel \textit{homography generation algorithm} that allows to continuously generate synthetic homographies at runtime, and by using SOTA DNNs.

\section{Methods}

\subsection{Theoretical Background}
Two images are related by a homography if both images view the same plane from different angles and distances. Points on the plane, as observed by the camera from different angles in homogeneous coordinates $\mathbf{p}_i = \begin{bmatrix}u_i&v_i&1\end{bmatrix}^\text{T}$ are related by a projective homography $\mathbf{G}$ \cite{malis2007deeper}
\begin{equation}
    \alpha_g\mathbf{p}_i = \mathbf{G}\mathbf{p}_i^\prime.
    \label{eq::proj_hom}
\end{equation}
Since the points $\mathbf{p}_i$ and $\mathbf{p}_i^\prime$ are only observed in the 2D image, depth information is lost, and the projective homography $\mathbf{G}$ can only be determined up to scale $\alpha_g$. The distinction between projective homography $\mathbf{G}$ and homography in Euclidean coordinates $\mathbf{H} = \mathbf{K}^{-1}\mathbf{G}\mathbf{K}$, with the camera intrinsics $\mathbf{K}$, is often not made for simplicity, but is nonetheless important for control purposes. The eight unknown parameters of $\mathbf{G}$ can be obtained through a set of $N\geq4$ matching points $\mathbb{P} = \{(\mathbf{p}_i, \mathbf{p}^\prime_i), i\in[0,N-1]\}$ by rearranging (\ref{eq::proj_hom}) into
\begin{equation}
    \begin{bmatrix}
        u^\prime_i & v^\prime_i & 1 & 0   &   0 & 0 & -u^\prime_i u_i & -v^\prime_i u_i & -u_i \\
        0   &   0 & 0 & u^\prime_i & v^\prime_i & 1 & -u^\prime_i v_i & -v^\prime_i v_i & - v_i
    \end{bmatrix}\mathbf{g}= \mathbf{0}\quad\forall i,
    \label{eq::hom_lin_sys}
\end{equation}
where $\mathbf{g}$ holds the entries of $\mathbf{G}$ as a column vector. The ninth constraint, by convention, is usually to set $||\mathbf{g}||_2 = 1$. Classically, $\mathbb{P}$ is obtained through feature detectors but it may also be used as a means to parameterise the spatial transformation.
Recent deep approaches indeed set $\mathbb{P}$ as the corners of an image, and predict $\Delta \mathbf{p}_i = \mathbf{p}^\prime_i - \mathbf{p}_i$. This is also known as the four point homography $\mathbf{G}_{4\text{point}}$
\begin{equation}
    \mathbf{G}_{4\text{point}} = \begin{bmatrix}
        \Delta u_0 & \Delta v_0 \\
        \Delta u_1 & \Delta v_1 \\
        \Delta u_2 & \Delta v_2 \\
        \Delta u_3 & \Delta v_3
    \end{bmatrix},
    \label{eq::4pt}
\end{equation}
which relates to $\mathbf{G}$ through (\ref{eq::hom_lin_sys}), where $\mathbf{p}^\prime_i = \mathbf{p}_i + \Delta \mathbf{p}_i$.    


\subsection{Data Preparation}

\begin{figure}[t!]
    \centering
    \subfloat[Da Vinci surgery datasets. Included are: \href{https://surgvisdom.grand-challenge.org/}{SurgVisDom} \cite{zia2021surgical}, \href{http://hamlyn.doc.ic.ac.uk/vision/}{GN} \cite{giannarou2012probabilistic}, \href{http://hamlyn.doc.ic.ac.uk/vision/}{MT} \cite{mountney2010three}, \href{https://saras-esad.grand-challenge.org/}{SARAS-ESAD} \cite{bawa2020esad}, \href{https://endovissub2017-kidneyboundarydetection.grand-challenge.org/}{KBD} \cite{hattab2020kidney}, \href{https://endovissub2017-roboticinstrumentsegmentation.grand-challenge.org/}{RIS} \cite{allan20192017}, \href{https://endovissub2018-roboticscenesegmentation.grand-challenge.org/home/}{RSS} \cite{allan20202018}.\label{fig::data_a}]{%
    \resizebox*{7cm}{!}{\includegraphics{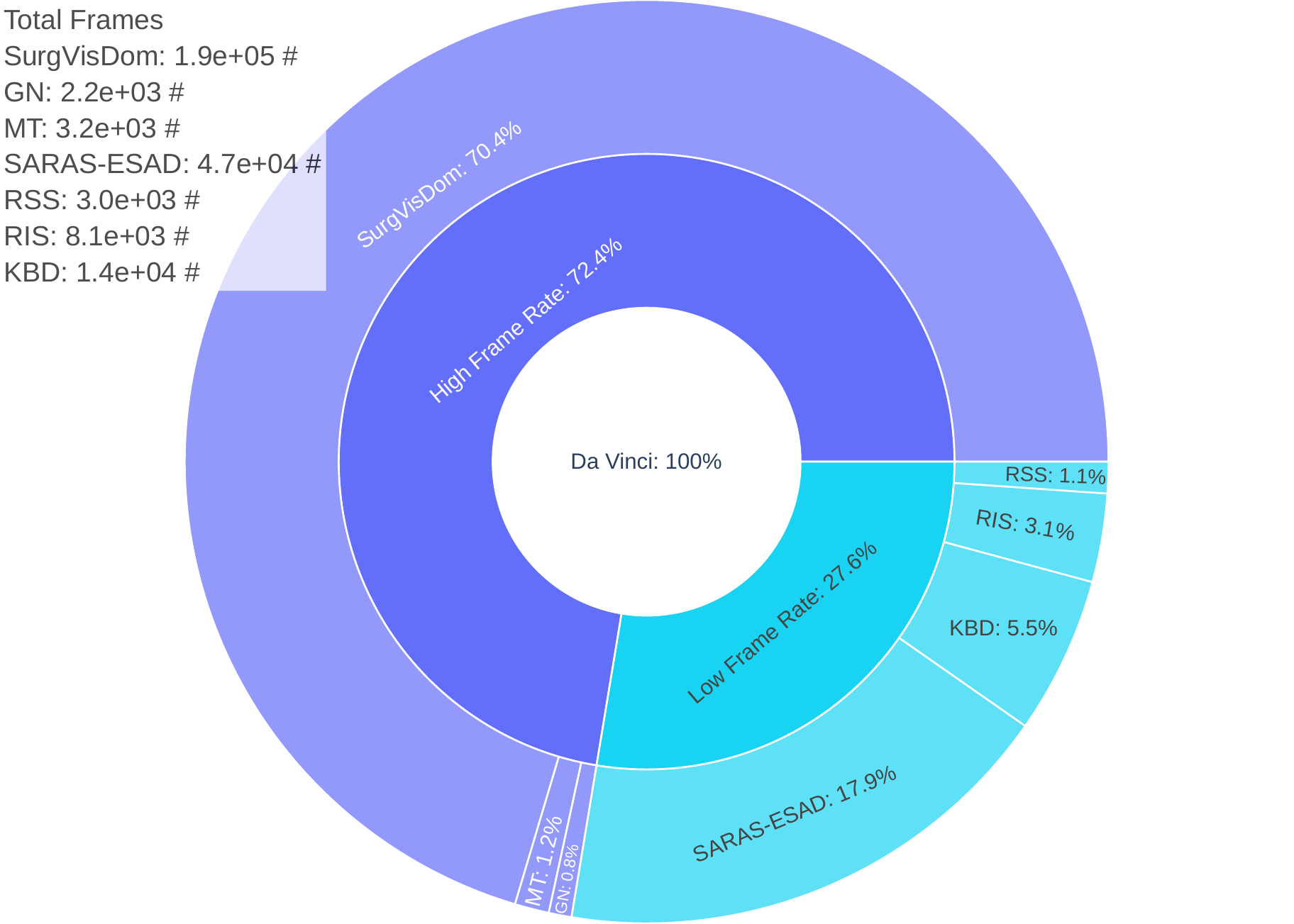}}}\hspace{5pt}
    \subfloat[Laparoscopic datasets and HFR da Vinci dataset from (a) for comparison. Included are: \href{https://robustmis2019.grand-challenge.org/}{ROBUST-MIS} \cite{maier2020heidelberg}, \href{http://camma.u-strasbg.fr/datasets}{Cholec80} \cite{twinanda2016endonet}, \href{https://endovissub-instrument.grand-challenge.org/}{ISAT} \cite{bodenstedt2018comparative}.\label{fig::data_b}]{%
    \resizebox*{7cm}{!}{\includegraphics{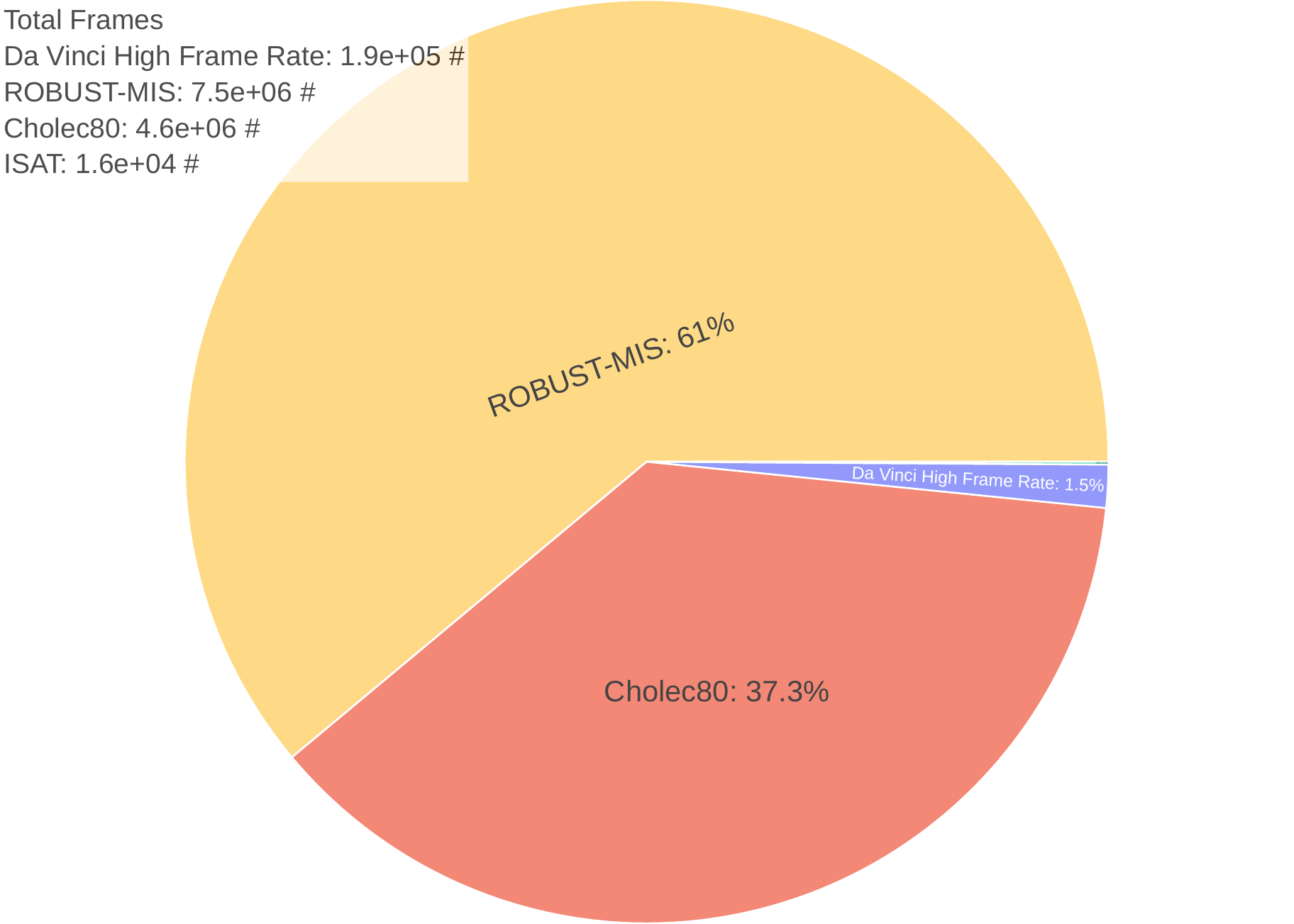}}}
    \caption{Da Vinci surgery and laparoscopic surgery datasets. Shown are relative sizes and the absolute number of frames. Da Vinci surgery datasets are often released at a low frame rate of $1\,\text{fps}$ for segmentation tasks (a). Much more laparoscopic surgery data is available (b).} \label{fig::data}
\end{figure}

Similar to \cite{le2020deep}, we initially find camera motion free image sequences, and synthetically add camera motion to them. In our work, we isolate camera motion free image sequences from da Vinci surgeries, and learn homography estimation supervisedly. We acquire publicly available laparoscopic, and da Vinci surgery videos. An overview of all datasets is shown in Fig.\,\ref{fig::data}. Excluded are synthetic, and publicly unavailable datasets. Da Vinci surgery datasets, and laparoscopic surgery datasets require different pre-processing steps, which are described below.

\subsubsection{Da Vinci Surgery Data Pre-Processing}

Many of the da Vinci surgery datasets are designed for tool or tissue segmentation tasks, therefore, they are published at a frame rate of $1\,\text{fps}$, see Fig.\,\ref{fig::data_a}. We merge all high frame rate (HFR) datasets into a single dataset and manually remove image sequences with camera motion, which amount to $5\%$ of all
HFR data. We crop the remaining data to remove status indicators, and scale the images to $306\times408$ pixels, later to be cropped by the \textit{homography generation algorithm} to a resolution of $240\times320$.

\subsubsection{Laparoscopic Surgery Data Pre-Processing}
\label{sec::lap_pre}

\begin{figure}[t!]
    \centering
    \subfloat[Binary segmentation mask, obtained through thresholding the bilateral filtered image.]{%
    \resizebox*{0.45\linewidth}{!}{\includegraphics{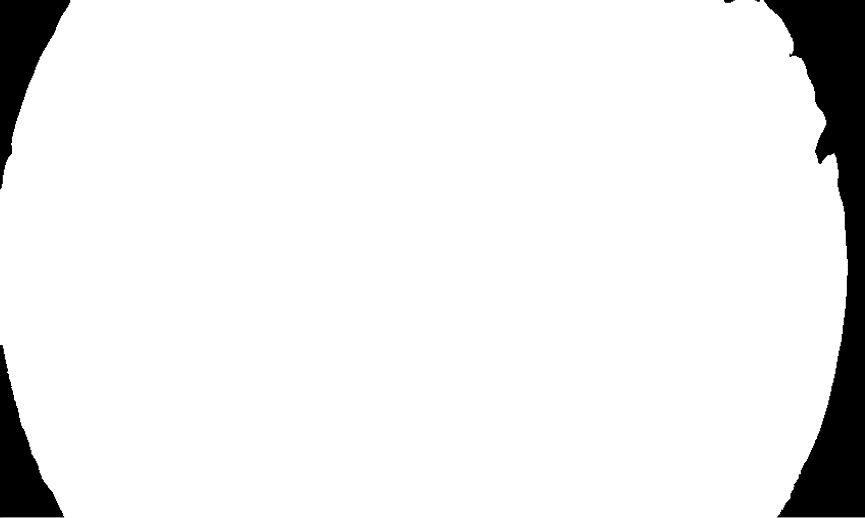}}}\hspace{0.08\linewidth}
    \subfloat[Circular boundary detection and static landmarks (blue arrows).\label{fig::seg_b}]{%
    \resizebox*{0.45\linewidth}{!}{\includegraphics{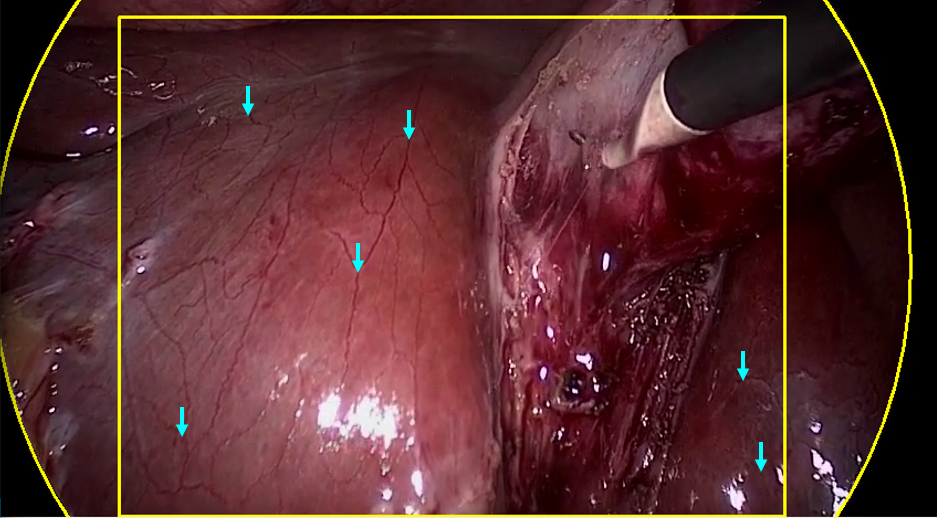}}}
    \caption{Cholec80 dataset pre-processing, referring to Sec.\,\ref{sec::lap_pre}. The black boundary circle is automatically detected. Landmarks are manually annotated and tracked over time (b).}
    \label{fig::seg}
\end{figure}

Laparoscopic images are typically observed through a Hopkins telescope, which causes a black circular boundary in the view, see Fig.\,\ref{fig::seg}. This boundary does not exist in da Vinci surgery recordings. For inference on the laparoscopic surgery image sequences, the most straightforward approach is to crop the view. To this purpose, we determine the center and radius of the circular boundary, which is only partially visible. We detect it by randomly sampling $N$ points $\mathbf{p}_i = (u_i,v_i)^\text{T}$ on the boundary. This is similar to work in \cite{munzer2013detection}, but instead of computing an analytical solution, we fit a circle by means of a least squares solution through inversion of
\begin{equation}
    \begin{bmatrix}
        2u_0     & 2v_0     & 1 \\
                 & \vdots   & \\
        2u_{N-1} & 2v_{N-1} & 1
    \end{bmatrix}
    \begin{bmatrix}
        x_0 \\ x_1 \\ x_2
    \end{bmatrix} = 
    \begin{bmatrix}
        u_0^2 + v_0^2 \\
        \vdots        \\
        u_{N-1}^2 + v_{N-1}^2
    \end{bmatrix},
\end{equation}
where the circle's center is $(x_0, x_1)$, and its radius is $\sqrt{x_2 + x^2_0 + x^2_1}$. We then crop the view centrally around the circle's center, and scale it to a resolution of $240\times320$. An implementation is provided on GitHub\footnote[2]{\url{https://github.com/RViMLab/endoscopy}}.

\subsubsection{Ground Truth Generation}
\label{sec::gt_gen}
One can simply use the synthetically generated camera motion as ground truth at train time. For inference on the laparoscopic dataset, this is not possible. We therefore generate ground truth data by randomly sampling $50$ image sequences with $10$ frames each from the Cholec80 dataset. In these image sequences, we find characteristic landmarks that are neither subject to tool, nor to object motion, see Fig.\,\ref{fig::seg_b}. Tracking of these landmarks over time allows one to estimate the camera motion in between consecutive frames through (\ref{eq::hom_lin_sys}).

\subsection{Deep Homography Estimation}

\begin{figure}[t!]
    \centering
    \includegraphics[width=\linewidth]{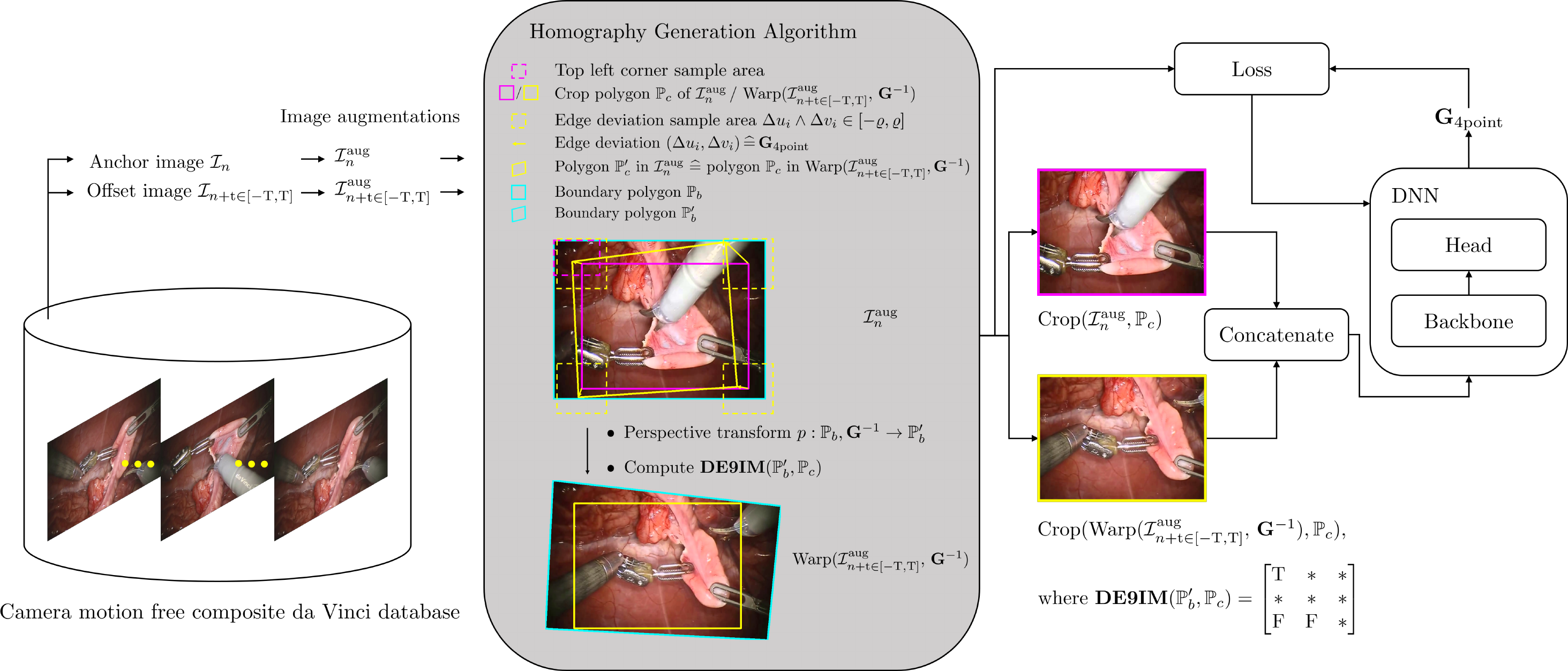}
    \caption{Deep homography estimation training pipeline. Image pairs are sampled from the HFR da Vinci surgery dataset. The \textit{homography generation algorithm} then adds synthetic camera motion to the augmented images, which is regressed through a backbone DNN.}
    \label{fig::hom}
\end{figure}

In this work we exploit the static camera in da Vinci surgeries, which allows us to isolate camera motion free image sequences. The processing pipeline is shown in Fig.\,\ref{fig::hom}.

Image pairs are sampled from image sequences of the HFR da Vinci surgery dataset of Fig.\,\ref{fig::data_a}. An image pair consists of an anchor image $\mathcal{I}_n$, and an offset image $\mathcal{I}_{n+t}$. The offset image is sampled uniformly from and interval $t\in[-T,T]$ around the anchor. The HFR da Vinci surgery dataset is relatively small, compared to the laparoscopic datasets, see Fig.\,\ref{fig::data_b}. Therefore, we apply image augmentations to the sampled image pairs. They include transform to grayscale, horizontal, and vertical flipping, cropping, change in brightness, and contrast, Gaussian blur, fog simulation, and random combinations of those. Camera motion is then added synthetically to the augmented image $\mathcal{I}^\text{aug}_{n+t}$ via the \textit{homography generation algorithm} from Sec.\,\ref{sec::hom_gen}. A DNN, with a backbone, then learns to predict the homography $\mathbf{G}_{4\text{point}}$ between the augmented image, and the augmented image with synthetic camera motion at time step $n+t$.

\subsection{Homography Generation Algorithm}
\label{sec::hom_gen}

In its core, the \textit{homography generation algorithm} is based on the works of \cite{detone2016deep}. However, where \citeauthor{detone2016deep} crop the image with a safety margin, our method allows to sample image crops across the entire image. Additionally, our method computes feasible homographies at runtime. This allows us to continuously generate synthetic camera motion, rather then training on a fixed set of precomputed homographies. The \textit{homography generation algorithm} is summarized in Alg.\,\ref{alg::hom}, and visualized in Fig.\,\ref{fig::hom}.

Initially, a \textit{crop polygon} $\mathbb{P}_c$ is generated for the augmented image $\mathcal{I}^\text{aug}_n$. The \textit{crop polygon} is defined through a set of points in the augmented image $\mathbb{P}_c = \{\mathbf{p}^c_i,\,i\in \left[0,3\right]\}$, which span a rectangle. The top left corner $\mathbf{p}^c_0$ is randomly sampled such that the \textit{crop polygon} $\mathbb{P}_c$ resides within the image border polygon $\mathbb{P}_b$, hence $\mathbf{p}^c_0 \in ([0, h_b - h_c], [0, w_b - w_c])$, where $h$, and $w$ are the height and width of the \textit{crop}, and the \textit{border polygon}, respectively. Following that, a random four point homography $\mathbf{G}_{4\text{point}}$ (\ref{eq::4pt}) is generated by sampling edge deviations $\Delta u_i \land \Delta v_i \in [-\varrho, \varrho]$. The corresponding inverse homography $\mathbf{G}^{-1}$ is used to warp each point of the border polygon $\mathbb{P}_b$ to $\mathbb{P}^\prime_b$. Finally, the Dimensionally Extended 9-Intersection Model \cite{clementini1994modelling} is used to determine whether the warped polygon $\mathbb{P}^\prime_b$ contains $\mathbb{P}_c$, for which we utilize the Python library \textit{Shapely}\footnote[3]{\url{https://pypi.org/project/Shapely}}. If the thus found intersection matrix $\IM$ satisfies
\begin{equation}
    \IM(\mathbb{P}^\prime_b, \mathbb{P}_c) = \begin{bmatrix}
        T & * & * \\
        * & * & * \\
        F & F & *
    \end{bmatrix},
\end{equation}
the homography $\mathbf{G}^{-1}$ is returned, otherwise a new four point homograpy $\mathbf{G}_{4\text{point}}$ is sampled. Therein, $*$ indicates that the intersection matrix may hold any value, and $T,\,F$ indicate that the intersection matrix must be true or false at the respective position. In the unlikely case that no homography is found after \textit{maximum rollouts}, the identity $\mathbf{G}_{4\text{point}} = \mathbf{0}$ is returned. Once a suitable homography is found, a crop of the augmented image $\Crop(\mathcal{I}^\text{aug}_n, \mathbb{P}_c)$ is computed, as well as a crop of the warped augmented image at time $n+t$, $\Crop(\Warp(\mathcal{I}^\text{aug}_{n+t}, \mathbf{G}^{-1}), \mathbb{P}_c)$. This keeps all computationally expensive operations outside the loop.

\begin{algorithm}
    \SetAlgoLined
    Randomly sample crop polygon $\mathbb{P}_c$ of desired shape in $\mathbb{P}_b$\;
    \While{rollouts $<$ maximum rollouts}{
        Randomly sample $\mathbf{G}_{4\text{point}}$, where $\Delta u_i \land \Delta v_i \in \left[-\varrho, \varrho\right] \forall i$\;
        Perspective transform boundary polygon $p: \mathbb{P}_b, \mathbf{G}^{-1}\rightarrow\mathbb{P}^\prime_b$\;
        Compute intersection matrix $\IM(\mathbb{P}^\prime_b, \mathbb{P}_c)$\;
        \If{$\IM = \begin{bmatrix}T & * & * \\ * & * & * \\ F & F & *\end{bmatrix}$}{
            \Return{$\mathbf{G}_{4\text{point}}$, $\mathbb{P}_c$}\;
        }
        Increment \textit{rollouts}\;
    }
    \Return{$\mathbf{0}$, $\mathbb{P}_c$}\;
    \caption{Homography generation algorithm.}
    \label{alg::hom}
\end{algorithm}

\section{Experiments}
We train DNNs on a $80\%$ train split of the HFR da Vinci surgery dataset from Fig.\,\ref{fig::data_a}. The $20\%$ test split is referred to as test set in the following. Inference is performed on the ground truth set from Sec.\,\ref{sec::gt_gen}. We compute the Mean Pairwise Distance (MPD) of the predicted value for $\mathbf{G}_{4\text{point}}$ from the desired one. We then compute the Cumulative Distribution Function (CDF) of all MPDs. We evaluate the CDF at different thresholds $t_i,\,i\in\{30,50,70.90\}$, e.g. $30\%$ of all homography estimations are below a MPD of $t_{30}$. We additionally evaluate the compute time on a GeForce RTX 2070 GPU, and a Intel Core i7-9750H CPU.

\subsection{Backbone Search}
In this experiment, we aim to find the best performing backbone for homography estimation. Therefore, we run the same experiment repeatedly with fixed hyperparameters, and varying backbones. We train each network for $50$ epochs, with a batch size of $64$, using the Adam optimizer with a learning rate of $\num{2e-4}$. The edge devation $\varrho$ is set to $32$, and the sequence length $T$ to $25$.

\subsection{Homography Generation Algorithm}
In this experiment, we evaluate the \textit{homography generation algorithm}. For this experiment we fix the backbone to a ResNet-34, and train it for $100$ epochs, with a batch size of $256$, using the Adam optimizer with a learning rate of \num{1e-3}. Initially, we fix the sequence length $T$ to $25$, and train on different edge deviations $\varrho\in\{32,48,64\}$. Next, we fix the edge deviation $\varrho$ to $48$, and train on different sequence lengths $T\in\{1,25,50\}$, where a sequence length of $1$ corresponds to a static pair of images.


\section{Results}




\subsection{Backbone Search}
\label{sec::backbone_search}



The results are listed in Tab.\,\ref{tab::results}. It can be seen that the deep methods generally outperform the classical methods on the test set. There is a tendency that models with more parameters perform better. On the ground truth set, this tendency vanishes. The differences in performance become independent of the number of parameters. Noticeably, many backbones still outperform the classical methods across all thresholds on the ground truth set, and low compute regime models also run quicker on CPU than comparable classical methods. E.g. we find that EfficientNet-B0, and RegNetY-400MF run at $36\,\text{Hz}$, and $50\,\text{Hz}$ on a CPU, respectively. Both outperform SURF \& RANSAC in homography estimation, which runs at $20\,\text{Hz}$.


\begin{table}[t!]
\centering
\tbl{Results referring to Sec.\,\ref{sec::backbone_search}. All methods are tested on the da Vinci HFR test set, indicated by $t^\text{test}_i$, and the Cholec80 inference set, indicated by $t^\text{gt}_i$. Best, and second best metrics are highlighted with bold character. Improvements in precision $t^\text{gt}_{90,\text{imp}}$ and compute time $\text{CPU}_\text{imp}$ are given w.r.t. SURF \& RANSAC. \label{tab::results}}{
    \resizebox{\linewidth}{!}{
        \begin{tabular}{lrrrrrrrrrr} \toprule
            Name            & $t^\text{test}_{30}/t^\text{gt}_{30}\,[\text{pixels}]$ & $t^\text{test}_{50}/t^\text{gt}_{50}\,[\text{pixels}]$ & $t^\text{test}_{70}/t^\text{gt}_{70}\,[\text{pixels}]$ & $t^\text{test}_{90}/t^\text{gt}_{90}\,[\text{pixels}]$ & $t^\text{gt}_{90,\text{imp}}\,[\%]$ & $\text{params}\,[\text{M}]$ & $\text{flops}\,[\text{M}]$ & $\text{GPU}\,[\text{ms}]$ & $\text{CPU}\,[\text{ms}]$ & $\text{CPU}_\text{imp}\,[\%]$ \\ \midrule
            VGG-style       & $4.83/2.45         $ & $ 6.47/2.94         $ & $ 8.68/3.59         $ & $ 13.23/5.41                  $ & $- 60         $ & $92.92$ & $11.12$ & $ \mathbf{2} \pm 1$ & $83          \pm 2$ & $- 69          \pm 33$ \\
            ResNet-18       & $1.42/1.12         $ & $ 1.95/1.33         $ & $ 2.82/1.58         $ & $  5.06/2.20                  $ & $  35         $ & $11.19$ & $ 6.02$ & $ \mathbf{3} \pm 1$ & $31          \pm 3$ & $  38          \pm 13$ \\
            ResNet-34       & $1.33/\mathbf{1.02}$ & $ 1.81/\mathbf{1.19}$ & $ 2.56/\mathbf{1.52}$ & $  4.63/2.08                  $ & $  \mathbf{39}$ & $21.3 $ & $11.74$ & $ 6          \pm 1$ & $51          \pm 5$ & $-  3          \pm 23$ \\
            ResNet-50       & $1.40/1.08         $ & $ 1.89/1.33         $ & $ 2.70/1.57         $ & $  4.79/2.21                  $ & $  35         $ & $23.53$ & $13.12$ & $10          \pm 1$ & $72          \pm 4$ & $- 46          \pm 29$ \\
            EfficientNet-B0 & $1.36/1.09         $ & $ 1.83/1.31         $ & $ 2.62/\mathbf{1.50}$ & $  4.64/\mathbf{2.01}         $ & $  \mathbf{41}$ & $ 4.02$ & $ 1.28$ & $12          \pm 2$ & $28          \pm 2$ & $  43          \pm 12$ \\
            EfficientNet-B1 & $1.32/\mathbf{1.02}$ & $ 1.77/\mathbf{1.26}$ & $ 2.50/1.57         $ & $  4.42/\mathbf{2.01}         $ & $  \mathbf{41}$ & $ 6.52$ & $ 1.88$ & $17          \pm 1$ & $37          \pm 1$ & $  25          \pm 15$ \\
            EfficientNet-B2 & $1.40/1.06         $ & $ 1.85/1.29         $ & $ 2.57/1.55         $ & $  4.42/2.15                  $ & $  37         $ & $ 7.71$ & $ 2.16$ & $17          \pm 2$ & $41          \pm 1$ & $  18          \pm 16$ \\
            EfficientNet-B3 & $1.31/1.05         $ & $ 1.75/1.36         $ & $ 2.44/1.68         $ & $  4.23/2.26                  $ & $  33         $ & $10.71$ & $ 3.14$ & $20          \pm 2$ & $55          \pm 4$ & $- 11          \pm 23$ \\
            EfficientNet-B4 & $\mathbf{1.23}/1.08$ & $ \mathbf{1.65}/1.31$ & $ \mathbf{2.29}/1.69$ & $  \mathbf{4.02}/2.14         $ & $  37         $ & $17.56$ & $ 4.88$ & $24          \pm 2$ & $68          \pm 5$ & $- 38          \pm 29$ \\
            EfficientNet-B5 & $1.26/1.18         $ & $ 1.67/1.35         $ & $ \mathbf{2.30}/1.65$ & $  \mathbf{4.02}/\mathbf{2.06}$ & $  \mathbf{39}$ & $28.36$ & $ 7.62$ & $29          \pm 2$ & $93          \pm 5$ & $- 89          \pm 37$ \\
            RegNetY-400MF   & $1.55/\mathbf{1.01}$ & $ 2.07/1.29         $ & $ 2.90/1.60         $ & $  5.08/2.12                  $ & $  37         $ & $ 3.91$ & $ 1.32$ & $13          \pm 1$ & $\mathbf{20} \pm 1$ & $  \mathbf{58} \pm  8$ \\
            RegNetY-600MF   & $1.47/1.03         $ & $ 1.98/1.28         $ & $ 2.80/1.56         $ & $  4.87/2.21                  $ & $  35         $ & $ 5.45$ & $ 1.94$ & $13          \pm 1$ & $24          \pm 3$ & $  52          \pm 11$ \\
            RegNetY-800MF   & $1.43/1.08         $ & $ 1.92/1.32         $ & $ 2.70/1.59         $ & $  4.76/2.12                  $ & $  37         $ & $ 5.50$ & $ 2.54$ & $12          \pm 1$ & $24          \pm 1$ & $  51          \pm 10$ \\
            RegNetY-1.6GF   & $1.38/1.03         $ & $ 1.83/1.27         $ & $ 2.52/1.60         $ & $  4.35/2.16                  $ & $  36         $ & $10.32$ & $ 5.08$ & $21          \pm 2$ & $42          \pm 4$ & $  16          \pm 18$ \\
            RegNetY-4.0GF   & $1.27/1.05         $ & $ 1.69/\mathbf{1.26}$ & $ 2.36/1.66         $ & $  4.17/2.17                  $ & $  36         $ & $19.57$ & $12.36$ & $21          \pm 2$ & $66          \pm 5$ & $- 34          \pm 28$ \\
            RegNetY-6.4GF   & $\mathbf{1.21}/1.04$ & $ \mathbf{1.64}/1.27$ & $ 2.32/1.60         $ & $  4.17/2.09                  $ & $  38         $ & $29.30$ & $19.72$ & $25          \pm 3$ & $98          \pm 6$ & $-100          \pm 40$ \\
            SURF \& RANSAC  & $4.06/1.07         $ & $ 5.65/1.40         $ & $ 7.93/2.02         $ & $ 13.62/3.39                  $ & $   0         $ & N/A     & N/A     &  N/A                & $49          \pm 9$ & $   0          \pm 27$ \\
            SIFT \& RANSAC  & $4.28/1.25         $ & $ 6.02/1.76         $ & $ 8.65/2.48         $ & $ 16.52/4.63                  $ & $- 37         $ & N/A     & N/A     &  N/A                & $37          \pm 9$ & $  25          \pm 22$ \\
            ORB \& RANSAC   & $6.52/1.65         $ & $10.48/2.47         $ & $20.12/3.71         $ & $122.66/6.81                  $ & $-101         $ & N/A     & N/A     &  N/A                & $\mathbf{12} \pm 2$ & $  \mathbf{76} \pm  6$ \\ \bottomrule
        \end{tabular}
    }
}
\end{table}
\subsection{Homography Generation Algorithm}
\label{sec::hom_opt}


\begin{figure}[t!]
    \centering
    \subfloat[Varying sequence length $T\in\{1,25,50\}$, and fixed edge deviation $\varrho=48$.\label{fig::resnet34_a}]{
        \resizebox{0.4\linewidth}{!}{\includegraphics{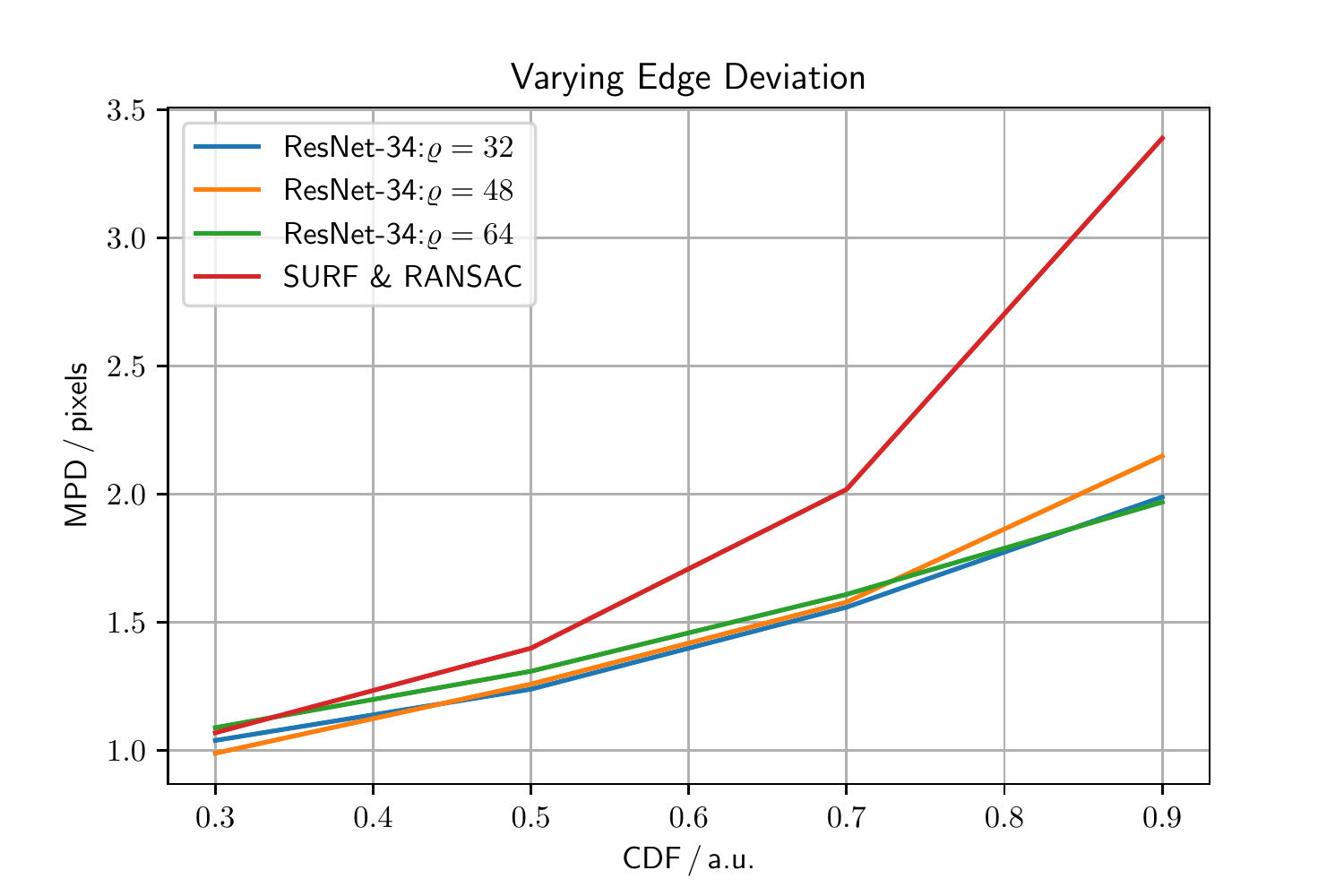}}
    }\hspace{0.04\linewidth}
    \subfloat[Varying edge deviation $\varrho\in\{32,48,64\}$, and fixed sequence length $T=25$.\label{fig::resnet34_b}]{
        \resizebox{0.4\linewidth}{!}{\includegraphics{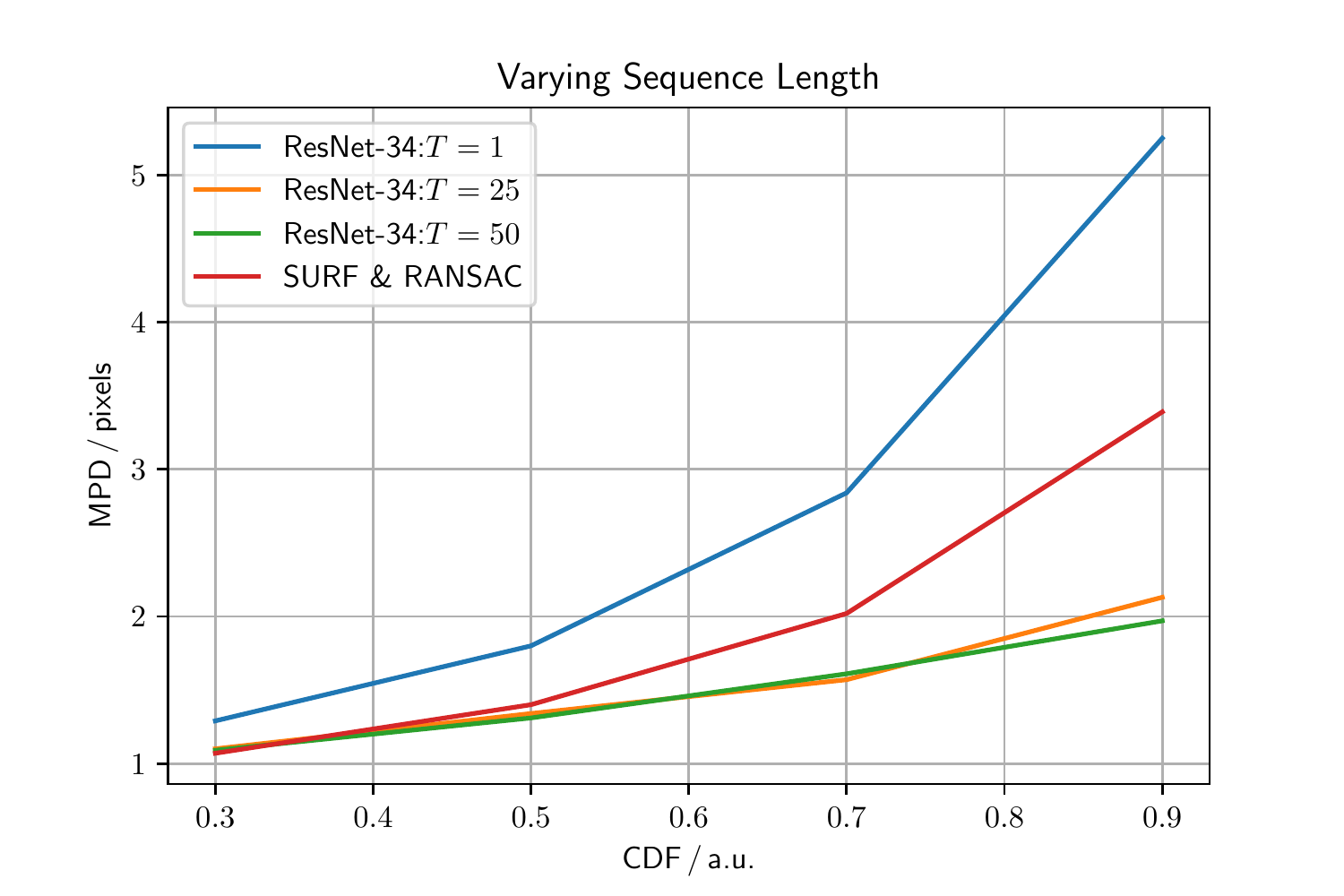}}
    }
    \caption{Homography generation optimization, referring to Sec.\,\ref{sec::hom_opt}. Shown is a ResNet-34 homography estimation for different homography generation configurations, and a SURF \& RANSAC homography estimation for reference. The edge deviation $\varrho$ is varied in (a), and the sequence length $T$ is varied in (b).}
    \label{fig::resnet34}
\end{figure}

\begin{figure}
    \centering
    \resizebox{0.6\linewidth}{!}{
        \includegraphics{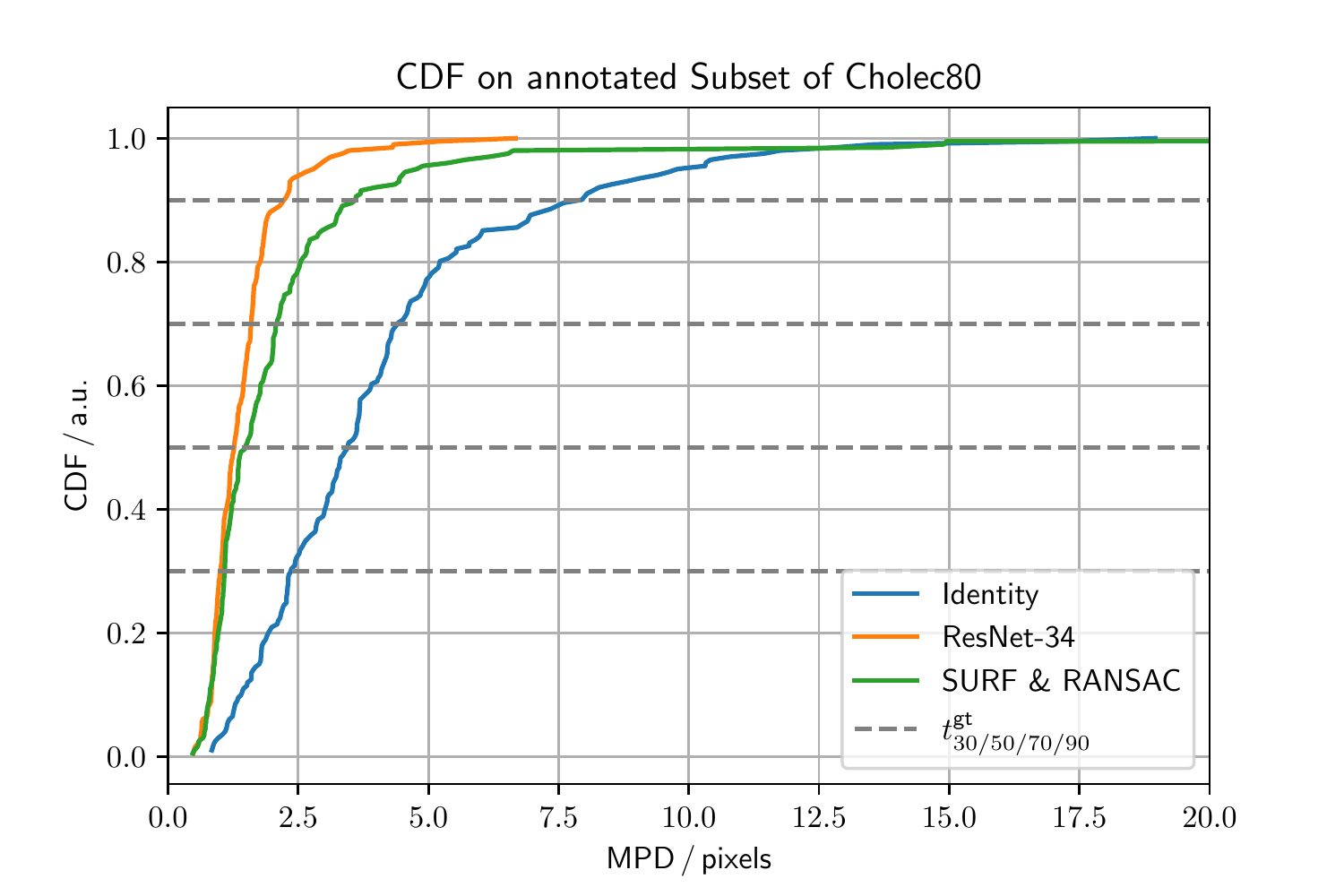}
    }
    \caption{CDF for SURF \& RANSAC, and ResNet-34, trained with a sequence length $T=25$, and edge deviation $\varrho=48$. The identity is added for reference. CDF thresholds for the SURF \& RANSAC are $t^\text{gt}_{1/10/30/50/70/90} = 0.51/0.80/1.09/1.48/2.07/3.53\,\text{pixels}$, and for the ResNet-34 $t^\text{gt}_{1/10/30/50/70/90} = 0.50/0.83/1.00/1.26/1.59/2.15\,\text{pixels}$. ResNet-34 generally performs better, and has no outliers.}
    \label{fig::resnet34_c}
\end{figure}

\begin{figure}
    \centering
    \subfloat[Identity]{%
    \resizebox*{0.25\linewidth}{!}{\includegraphics{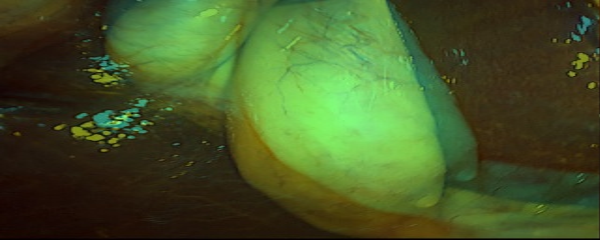}}}\hspace{0.06\linewidth}
    \subfloat[SURF \& RANSAC]{%
    \resizebox*{0.25\linewidth}{!}{\includegraphics{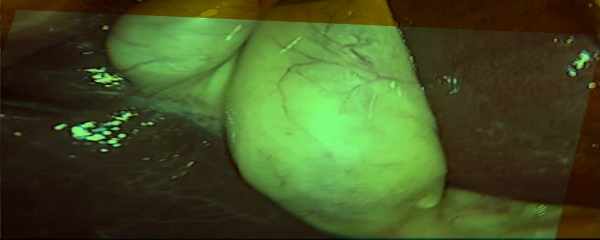}}}\hspace{0.06\linewidth}
    \subfloat[ResNet-34]{%
    \resizebox*{0.25\linewidth}{!}{\includegraphics{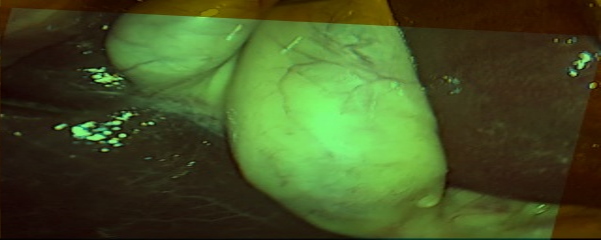}}}
    \vspace{-0.01\linewidth}
    \subfloat[Object Motion - SURF \& RANSAC, zoom]{%
    \resizebox*{0.363125\linewidth}{!}{\includegraphics{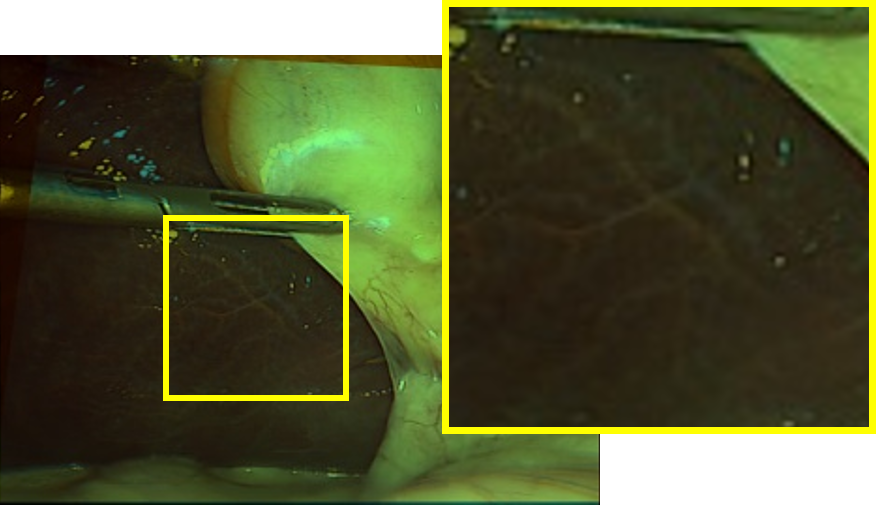}}}\hspace{0.155\linewidth}
    \subfloat[Object Motion - ResNet-34, zoom]{%
    \resizebox*{0.363125\linewidth}{!}{\includegraphics{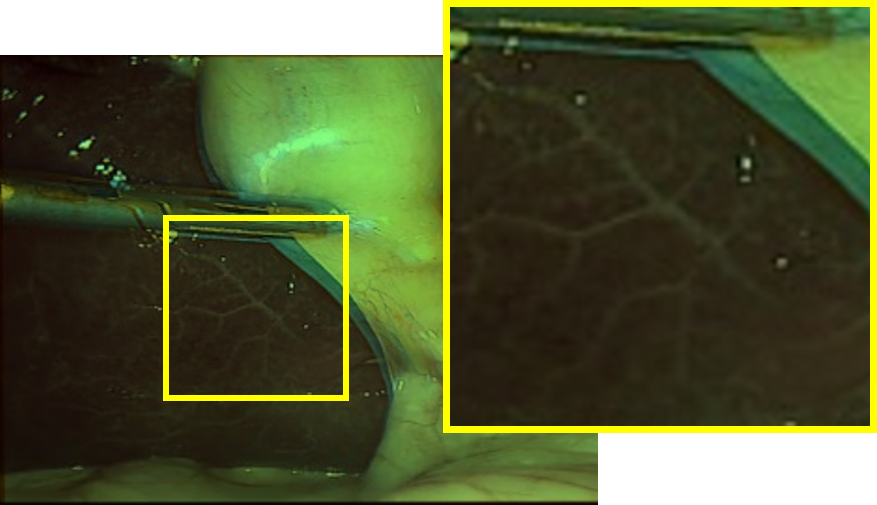}}}
    \caption{Classical homography estimation using a SURF feature detector under RANSAC outlier rejection, and the proposed deep homography estimation with a ResNet-34 backbone, referring to Sec.\,\ref{sec::hom_opt}. Shown are blends of consecutive images from a $5\,\text{fps}$ resampled Cholec80 exemplary sequence \cite{twinanda2016endonet}. Decreasing the framerate from originally $25\,\text{fps}$ to $5\,\text{fps}$, increases the motion in between consecutive frames. (Top row) Homography estimation under predominantly camera motion. Both methods perform well. (Bottom row) Homography estimation under predominantly object motion. Especially in the zoomed images it can be seen that the classical method (d) misaligns the stationary parts of the image, whereas the proposed method (e) aligns the background well.}
    \label{fig::qualitative}
\end{figure}

Given that ResNet-34 performs well on the ground truth set, and executes fast on the GPU, we run the \textit{homography generation algorithm} experiments with it. It can be seen in Fig.\,\ref{fig::resnet34_a}, that the edge deviation $\varrho$ is neglectable for inference. In Fig.\,\ref{fig::resnet34_b}, one sees the effects of the sequence length $T$ on the inference performance. Notably, with $T=1$, corresponding to static image pairs, the SURF \& RANSAC homography estimation outperforms the ResNet-34. For the other sequence lengths, ResNet-34 outperforms the classical homography estimation. The CDF for the best performing combination of parameters, with $T=25$, and $\varrho=48$, is shown in Fig.\,\ref{fig::resnet34_c}. Our method generally outperforms SURF \& RANSAC. The advantage of our method becomes most apparent for a $\text{CDF}\geq0.5$. Even the identity outperforms SURF \& RANSAC for large MPDs. This aligns with the qualitative observation that motion is often overestimated by SURF \& RANSAC, which is shown in Fig.\,\ref{fig::qualitative}. An exemplary video is provided\footnote{\url{https://drive.google.com/file/d/1totjHbhIMEL7a-QAiL7B1rT44wvWB6lO/view?usp=sharing}}.




\section{Conclusion}

In this work we supervisedly learn homography estimation in dynamic surgical scenes. We train our method on a newly acquired, synthetically modified da Vinci surgery dataset and successfully cross the domain gap to videos of laparoscopic surgeries. To do so, we introduce extensive data augmentation and continuously generate synthetic camera motion through a novel \textit{homography generation algorithm}.

In Sec.\,\ref{sec::backbone_search}, we find that, despite the domain gap for the ground truth set, DNNs outperform classical methods, which is indicated in Tab.\,\ref{tab::results}. The homography estimation performance proofs to be independent of the number of model parameters, which indicates an overfit to the test data. The independence of the number of parameters allows to optimize the backbone for computational requirements. E.g., a typical laparoscopic setup runs at $25-30\,\text{Hz}$, the classical method would thus already introduce a bottleneck at $20\,\text{Hz}$. On the other hand, EfficientNet-B0, with $36\,\text{Hz}$, and RegNetY-400MF, with $50\,\text{Hz}$, introduce no latency, and could be integrated into systems without GPU.


In Sec.\,\ref{sec::hom_opt}, we find that increasing the edge deviation has no effect on the homography estimation, see Fig.\,\ref{fig::resnet34_a}. This is because the motion in the ground truth set does not exceed the motion in the training set. In Fig.\,\ref{fig::resnet34_b}, we further find how training DNNs on synthetically modified da Vinci surgery image sequences enables our method to isolate camera from object and tool motion, validating our method. In Fig.\,\ref{fig::resnet34_c}, it is demonstrated that ResNet-34 generally outperforms SURF \& RANSAC. This shows that generating camera motion synthetically through homographies, which approximates the surgical scene as a plane, does not pose an issue.


The object, and tool motion invariant camera motion estimation allows one to extract a laparoscope holder's actions from videos of laparoscopic interventions, which enables the generation of image-action-pairs. In future work, we will generate image-action-pairs from laparoscopic datasets and apply IL to them. Describing camera motion (actions) by means of a homography is grounded in recent research for robotic control of laparoscopes \cite{huber2021homographybased}. This work will therefore support the transition towards robotic automation approaches. It might further improve augmented reality, and image mosaicing methods in dynamic surgical environments.

\section*{Acknowledgements and Disclosures}

This work was supported by core and project funding from the Wellcome/EPSRC [WT203148/Z/16/Z; NS/A000049/1; WT101957; NS/A000027/1]. This project has received funding from the European Union's Horizon 2020 research and innovation programme under grant agreement No 101016985 (FAROS project).
TV is supported by a Medtronic / RAEng Research Chair [RCSRF1819\textbackslash7\textbackslash34].
SO and TV are co-founders and shareholders of Hypervision Surgical.
TV holds shares from Mauna Kea Technologies.


\bibliographystyle{tfcse}
\bibliography{mybib}

\begin{thebibliography}{39}
\providecommand{\natexlab}[1]{#1}
\providecommand{\url}[1]{\normalfont{#1}}
\providecommand{\urlprefix}{Available from: }

\bibitem[Agustinos et~al.(2014)]{agustinos2014visual}
Agustinos~A, Wolf~R, Long~JA, Cinquin~P, Voros~S. 2014. Visual servoing of a
  robotic endoscope holder based on surgical instrument tracking. In: 5th IEEE
  RAS/EMBS International Conference on Biomedical Robotics and Biomechatronics.
  IEEE. p. 13--18.

\bibitem[Allan et~al.(2020)]{allan20202018}
Allan~M, Kondo~S, Bodenstedt~S, Leger~S, Kadkhodamohammadi~R, Luengo~I,
  Fuentes~F, Flouty~E, Mohammed~A, Pedersen~M, et~al. 2020. 2018 robotic scene
  segmentation challenge. arXiv preprint arXiv:200111190.

\bibitem[Allan et~al.(2019)]{allan20192017}
Allan~M, Shvets~A, Kurmann~T, Zhang~Z, Duggal~R, Su~YH, Rieke~N, Laina~I,
  Kalavakonda~N, Bodenstedt~S, et~al. 2019. 2017 robotic instrument
  segmentation challenge. arXiv preprint arXiv:190206426.

\bibitem[Bano et~al.(2020)]{bano2020deep}
Bano~S, Vasconcelos~F, Tella-Amo~M, Dwyer~G, Gruijthuijsen~C, Vander~Poorten~E,
  Vercauteren~T, Ourselin~S, Deprest~J, Stoyanov~D. 2020. Deep learning-based
  fetoscopic mosaicking for field-of-view expansion. International journal of
  computer assisted radiology and surgery. 15(11):1807--1816.

\bibitem[Bawa et~al.(2020)]{bawa2020esad}
Bawa~VS, Singh~G, KapingA~F, Leporini~A, Landolfo~C, Stabile~A, Setti~F,
  Muradore~R, Oleari~E, Cuzzolin~F, et~al. 2020. Esad: Endoscopic surgeon
  action detection dataset. arXiv preprint arXiv:200607164.

\bibitem[Bodenstedt et~al.(2018)]{bodenstedt2018comparative}
Bodenstedt~S, Allan~M, Agustinos~A, Du~X, Garcia-Peraza-Herrera~L, Kenngott~H,
  Kurmann~T, M{\"u}ller-Stich~B, Ourselin~S, Pakhomov~D, et~al. 2018.
  Comparative evaluation of instrument segmentation and tracking methods in
  minimally invasive surgery. arXiv preprint arXiv:180502475.

\bibitem[Clementini et~al.(1994)]{clementini1994modelling}
Clementini~E, Sharma~J, Egenhofer~MJ. 1994. Modelling topological spatial
  relations: Strategies for query processing. Computers \& graphics.
  18(6):815--822.

\bibitem[Da~Col et~al.(2020)]{da2020scan}
Da~Col~T, Mariani~A, Deguet~A, Menciassi~A, Kazanzides~P, De~Momi~E. 2020.
  Scan: System for camera autonomous navigation in robotic-assisted surgery.
  In: 2020 IEEE/RSJ International Conference on Intelligent Robots and Systems
  (IROS). IEEE. p. 2996--3002.

\bibitem[DeTone et~al.(2016)]{detone2016deep}
DeTone~D, Malisiewicz~T, Rabinovich~A. 2016. Deep image homography estimation.
  arXiv preprint arXiv:160603798.

\bibitem[Ellis et~al.(2016)]{ellis2016task}
Ellis~RD, Munaco~AJ, Reisner~LA, Klein~MD, Composto~AM, Pandya~AK, King~BW.
  2016. Task analysis of laparoscopic camera control schemes. The International
  Journal of Medical Robotics and Computer Assisted Surgery. 12(4):576--584.

\bibitem[Erlik~Nowruzi et~al.(2017)]{erlik2017homography}
Erlik~Nowruzi~F, Laganiere~R, Japkowicz~N. 2017. Homography estimation from
  image pairs with hierarchical convolutional networks. In: Proceedings of the
  IEEE International Conference on Computer Vision Workshops. p. 913--920.

\bibitem[Esteva et~al.(2019)]{esteva2019guide}
Esteva~A, Robicquet~A, Ramsundar~B, Kuleshov~V, DePristo~M, Chou~K, Cui~C,
  Corrado~G, Thrun~S, Dean~J. 2019. A guide to deep learning in healthcare.
  Nature medicine. 25(1):24--29.

\bibitem[Garcia-Peraza-Herrera et~al.(2021)]{garcia2021image}
Garcia-Peraza-Herrera~LC, Fidon~L, D’Ettorre~C, Stoyanov~D, Vercauteren~T,
  Ourselin~S. 2021. Image compositing for segmentation of surgical tools
  without manual annotations. IEEE transactions on medical imaging.
  40(5):1450--1460.

\bibitem[Garcia-Peraza-Herrera et~al.(2017)]{garcia2017toolnet}
Garcia-Peraza-Herrera~LC, Li~W, Fidon~L, Gruijthuijsen~C, Devreker~A,
  Attilakos~G, Deprest~J, Vander~Poorten~E, Stoyanov~D, Vercauteren~T, et~al.
  2017. Toolnet: holistically-nested real-time segmentation of robotic surgical
  tools. In: 2017 IEEE/RSJ International Conference on Intelligent Robots and
  Systems (IROS). IEEE. p. 5717--5722.

\bibitem[Giannarou et~al.(2012)]{giannarou2012probabilistic}
Giannarou~S, Visentini-Scarzanella~M, Yang~GZ. 2012. Probabilistic tracking of
  affine-invariant anisotropic regions. IEEE transactions on pattern analysis
  and machine intelligence. 35(1):130--143.

\bibitem[Gomes et~al.(2019)]{gomes2019unsupervised}
Gomes~S, Val{\'e}rio~MT, Salgado~M, Oliveira~HP, Cunha~A. 2019. Unsupervised
  neural network for homography estimation in capsule endoscopy frames.
  Procedia Computer Science. 164:602--609.

\bibitem[Gruijthuijsen et~al.(2021)]{gruijthuijsen2021autonomous}
Gruijthuijsen~C, Garcia-Peraza-Herrera~LC, Borghesan~G, Reynaerts~D, Deprest~J,
  Ourselin~S, Vercauteren~T, Poorten~EV. 2021. Autonomous robotic endoscope
  control based on semantically rich instructions. arXiv preprint
  arXiv:210702317.

\bibitem[Hattab et~al.(2020)]{hattab2020kidney}
Hattab~G, Arnold~M, Strenger~L, Allan~M, Arsentjeva~D, Gold~O,
  Simpfend{\"o}rfer~T, Maier-Hein~L, Speidel~S. 2020. Kidney edge detection in
  laparoscopic image data for computer-assisted surgery. International journal
  of computer assisted radiology and surgery. 15(3):379--387.

\bibitem[Huber et~al.(2021)]{huber2021homographybased}
Huber~M, Mitchell~JB, Henry~R, Ourselin~S, Vercauteren~T, Bergeles~C. 2021.
  Homography-based visual servoing with remote center of motion for
  semi-autonomous robotic endoscope manipulation. arXiv preprint
  arXiv:211013245.

\bibitem[Ji et~al.(2018)]{ji2018learning}
Ji~JJ, Krishnan~S, Patel~V, Fer~D, Goldberg~K. 2018. Learning 2d surgical
  camera motion from demonstrations. In: 2018 IEEE 14th International
  Conference on Automation Science and Engineering (CASE). IEEE. p. 35--42.

\bibitem[Kassahun et~al.(2016)]{kassahun2016surgical}
Kassahun~Y, Yu~B, Tibebu~AT, Stoyanov~D, Giannarou~S, Metzen~JH,
  Vander~Poorten~E. 2016. Surgical robotics beyond enhanced dexterity
  instrumentation: a survey of machine learning techniques and their role in
  intelligent and autonomous surgical actions. International journal of
  computer assisted radiology and surgery. 11(4):553--568.

\bibitem[Kitaguchi et~al.(2020)]{kitaguchi2020real}
Kitaguchi~D, Takeshita~N, Matsuzaki~H, Takano~H, Owada~Y, Enomoto~T, Oda~T,
  Miura~H, Yamanashi~T, Watanabe~M, et~al. 2020. Real-time automatic surgical
  phase recognition in laparoscopic sigmoidectomy using the convolutional
  neural network-based deep learning approach. Surgical endoscopy.
  34(11):4924--4931.

\bibitem[Le et~al.(2020)]{le2020deep}
Le~H, Liu~F, Zhang~S, Agarwala~A. 2020. Deep homography estimation for dynamic
  scenes. In: Proceedings of the IEEE/CVF Conference on Computer Vision and
  Pattern Recognition. p. 7652--7661.

\bibitem[Maier-Hein et~al.(2020)]{maier2020heidelberg}
Maier-Hein~L, Wagner~M, Ross~T, Reinke~A, Bodenstedt~S, Full~PM, Hempe~H,
  Mindroc-Filimon~D, Scholz~P, Tran~TN, et~al. 2020. Heidelberg colorectal data
  set for surgical data science in the sensor operating room. arXiv preprint
  arXiv:200503501.

\bibitem[Malis and Vargas(2007)]{malis2007deeper}
Malis~E, Vargas~M. 2007. Deeper understanding of the homography decomposition
  for vision-based control [dissertation]. INRIA.

\bibitem[Mountney et~al.(2010)]{mountney2010three}
Mountney~P, Stoyanov~D, Yang~GZ. 2010. Three-dimensional tissue deformation
  recovery and tracking. IEEE Signal Processing Magazine. 27(4):14--24.

\bibitem[M{\"u}nzer et~al.(2013)]{munzer2013detection}
M{\"u}nzer~B, Schoeffmann~K, B{\"o}sz{\"o}rmenyi~L. 2013. Detection of circular
  content area in endoscopic videos. In: Proceedings of the 26th IEEE
  International Symposium on Computer-Based Medical Systems. IEEE. p. 534--536.

\bibitem[Nguyen et~al.(2018)]{nguyen2018unsupervised}
Nguyen~T, Chen~SW, Shivakumar~SS, Taylor~CJ, Kumar~V. 2018. Unsupervised deep
  homography: A fast and robust homography estimation model. IEEE Robotics and
  Automation Letters. 3(3):2346--2353.

\bibitem[Pandya et~al.(2014)]{pandya2014review}
Pandya~A, Reisner~LA, King~B, Lucas~N, Composto~A, Klein~M, Ellis~RD. 2014. A
  review of camera viewpoint automation in robotic and laparoscopic surgery.
  Robotics. 3(3):310--329.

\bibitem[Rivas-Blanco et~al.(2014)]{rivas2014towards}
Rivas-Blanco~I, Estebanez~B, Cuevas-Rodriguez~M, Bauzano~E, Mu{\~n}oz~VF. 2014.
  Towards a cognitive camera robotic assistant. In: 5th IEEE RAS/EMBS
  International Conference on Biomedical Robotics and Biomechatronics. IEEE. p.
  739--744.

\bibitem[Rivas-Blanco et~al.(2017)]{rivas2017smart}
Rivas-Blanco~I, L{\'o}pez-Casado~C, P{\'e}rez-del Pulgar~CJ, Garcia-Vacas~F,
  Fraile~J, Mu{\~n}oz~VF. 2017. Smart cable-driven camera robotic assistant.
  IEEE Transactions on Human-Machine Systems. 48(2):183--196.

\bibitem[Rivas-Blanco et~al.(2019)]{rivas2019transferring}
Rivas-Blanco~I, Perez-del Pulgar~CJ, L{\'o}pez-Casado~C, Bauzano~E,
  Mu{\~n}oz~VF. 2019. Transferring know-how for an autonomous camera robotic
  assistant. Electronics. 8(2):224.

\bibitem[Simonyan and Zisserman(2014)]{simonyan2014very}
Simonyan~K, Zisserman~A. 2014. Very deep convolutional networks for large-scale
  image recognition. arXiv preprint arXiv:14091556.

\bibitem[Su et~al.(2020)]{su2020multicamera}
Su~YH, Huang~K, Hannaford~B. 2020. Multicamera 3d reconstruction of dynamic
  surgical cavities: Autonomous optimal camera viewpoint adjustment. In: 2020
  International Symposium on Medical Robotics (ISMR). IEEE. p. 103--110.

\bibitem[Twinanda et~al.(2016)]{twinanda2016endonet}
Twinanda~AP, Shehata~S, Mutter~D, Marescaux~J, De~Mathelin~M, Padoy~N. 2016.
  Endonet: a deep architecture for recognition tasks on laparoscopic videos.
  IEEE transactions on medical imaging. 36(1):86--97.

\bibitem[Wagner et~al.(2021)]{wagner2021learning}
Wagner~M, Bihlmaier~A, Kenngott~HG, Mietkowski~P, Scheikl~PM, Bodenstedt~S,
  Schiepe-Tiska~A, Vetter~J, Nickel~F, Speidel~S, et~al. 2021. A learning robot
  for cognitive camera control in minimally invasive surgery. Surgical
  Endoscopy:1--10.

\bibitem[Weede et~al.(2011)]{weede2011intelligent}
Weede~O, M{\"o}nnich~H, M{\"u}ller~B, W{\"o}rn~H. 2011. An intelligent and
  autonomous endoscopic guidance system for minimally invasive surgery. In:
  2011 IEEE International Conference on Robotics and Automation. IEEE. p.
  5762--5768.

\bibitem[Zhang et~al.(2020)]{zhang2020content}
Zhang~J, Wang~C, Liu~S, Jia~L, Ye~N, Wang~J, Zhou~J, Sun~J. 2020. Content-aware
  unsupervised deep homography estimation. In: European Conference on Computer
  Vision. Springer. p. 653--669.

\bibitem[Zia et~al.(2021)]{zia2021surgical}
Zia~A, Bhattacharyya~K, Liu~X, Wang~Z, Kondo~S, Colleoni~E, van Amsterdam~B,
  Hussain~R, Hussain~R, Maier-Hein~L, et~al. 2021. Surgical visual domain
  adaptation: Results from the miccai 2020 surgvisdom challenge.

\end{thebibliography}


\appendix

\end{document}